\documentclass[12pt]{article}

\usepackage{amsfonts,amsmath,amssymb}
\usepackage{hyperref}
\usepackage{cite}
\usepackage{epsfig}
\usepackage{paralist}
\usepackage{fancyhdr}
\usepackage{tikz}
\usepackage{tkz-euclide}
\usetikzlibrary{decorations.pathmorphing,patterns,calc,snakes,arrows}

\usepackage{graphicx}
\usepackage{xcolor}
\numberwithin{equation}{section}
\usepackage[vcentermath]{youngtab}
\usepackage{etex}
\usepackage{braket}
\usepackage{float}

\def\spa#1{\phantom{\fbox{\rule[-#1cm]{0cm}{0cm}}}}

\def\be{\begin{equation}}
\def\ee{\end{equation}}
\def\bea{\begin{eqnarray}}
\def\eea{\end{eqnarray}}

\def\del{\partial}

\renewcommand{\thefootnote}{\fnsymbol{footnote}}

\makeatletter
\g@addto@macro\bfseries{\boldmath}
\makeatother

\begin{document}

\hfuzz=100pt
\title{{\Large \bf{Island Formula from Wald-like Entropy with Backreaction}}}
\date{}
\author{Shinji Hirano$^{a, b}$\footnote{
	e-mail:
	\href{mailto:shinji.hirano@wits.ac.za}{shinji.hirano@gmail.com}}
}
\date{}

\maketitle

\thispagestyle{fancy}
\rhead{YITP-23-127}
\cfoot{}
\renewcommand{\headrulewidth}{0.0pt}

\vspace*{-1cm}
\begin{center}
$^{a}${{\it School of Science, Huzhou University}}
\\ {{\it Huzhou 313000, Zhejiang, China}}
  \spa{0.5} \\
$^b${{\it Center for Gravitational Physics and Quantum Information (CGPQI)}}
\\ {{\it  Yukawa Institute for Theoretical Physics, Kyoto University}}
\\ {{\it Kitashirakawa-Oiwakecho, Sakyo-ku, Kyoto 606-8502, Japan}}
\spa{0.5}  

\end{center}

\begin{abstract}
We propose a Lorentzian derivation of the generalized entropy associated with the island formula for black holes as a Wald-like entropy without reference to the exterior non-gravitating region or field-theoretic von Neumann entropy of Hawking radiation in a fixed curved spacetime background. We illustrate this idea by studying two-dimensional black holes in the Jackiw-Teitelboim gravity and the Russo-Susskind-Thorlacius model in which Hawking radiation is represented by conformal scalars. With some prescriptions assumed, we show that the generalized entropy for the island formula can be reproduced as the Wald-like entropy of the two-dimensional dilaton-gravity theories upon the inclusion of the backreaction from Hawking radiation described by conformal anomaly. We give a discussion on how a similar idea can be applied to higher-dimensional black holes.
It is emphasized that the generalized entropy is obtained in a fully gravitational fashion, yet it yields the same Page curve as that of the half-gravitational set-up. We argue that the results in this paper exacerbate the issues raised in the work of massive islands and inconsistency of islands in theories of long-range gravity.

\end{abstract}

\renewcommand{\thefootnote}{\arabic{footnote}}
\setcounter{footnote}{0}

\newpage

\tableofcontents


\section{Introduction}
\label{Sec:Introduction}

Since the discovery of Hawking radiation \cite{Hawking:1975vcx}, the black hole information paradox is considered to be one of the most outstanding problems in understanding the consistency of general relativity and quantum mechanics. Tantalizingly, the black hole behaves as a thermodynamical system \cite{Bekenstein:1972tm, Bekenstein:1973ur, Bardeen:1973gs, Hawking:1975vcx} and as a consequence, it is as if the information of the pure states that created or fell into the black hole were lost into a mixed thermal state. 
Over the last few years, there has been a potentially important development towards the resolution of the information paradox that is often colloquially called the island proposal \cite{Almheiri:2020cfm}. Evidence for the proposal is the so-called Page curve for the radiation entropy  \cite{Page:1993wv, Page:2013dx}, indicative of the unitary evolution, reproduced by the island formula we are now going to review \cite{Penington:2019npb,Almheiri:2019psf, Almheiri:2019hni}.

The island formula \cite{Penington:2019npb,Almheiri:2019psf, Almheiri:2019hni, Penington:2019kki,Almheiri:2019qdq } for black holes is built on the (semi-classical) generalized entropy $S_{\rm gen}$ \cite{Bekenstein:1973ur}. The claim is that the von Neumann entropy $S$ of the black hole, or equivalently, for pure states, of Hawking radiation is given by\footnote{The idea of the island formula is nothing holographic and applicable to gravitational theories without field theory duals but was inspired by the development in understanding the gravitational von Neumann (or fine-grained) entropy in the AdS/CFT correspondence \cite{Ryu:2006bv, Hubeny:2007xt, Lewkowycz:2013nqa, Faulkner:2013ana,Engelhardt:2014gca}.}
\begin{align}\label{islandformula}
S={\rm min}_X\biggl[{\rm ext}_X S_{\rm gen}(X, \Sigma_X)\biggr]
\qquad\mbox{where}\qquad S_{\rm gen}={{\rm Area}(X)\over 4G_N}+S_{\rm vN}(\Sigma_X)\ .
\end{align}
A crucial part in this proposal is that $S$ on the LHS is the von Neumann entropy of the full quantum state of radiation, whereas $S_{\rm vN}$ on the RHS is that in the semi-classical description. It is a surprise and somewhat of mystery that a semi-classical computation gives the full quantum result. To explain more details of the formula, $X$ is a codimension-two surface, which is a certain generalization of the black hole horizon, and $\Sigma_X$ is a spacelike region $B_X$ bounded by $X$ and an outer surface $D$ collecting the Hawking radiation, or equivalently, for pure states, its complement $\bar{B}_X=I_X\cup R$. The island $I_X$ is the spacelike region bounded by $X=\partial I_X$ and extending into the interior of the black hole when the extremal surface $X$ exists. The radiation region $R$ is the spacelike region bounded by $D=\partial R$ and extending to the asymptotic end of the spacetime. The surface $X$ needs to be extremized and the generalized entropy $S_{\rm gen}$ is then minimized with respect to all the extremal surfaces \cite{Wall:2012uf, Akers:2019lzs}. Note that the empty set $X=\O$ (no-island) is included as a part of the definition of extrema. 
At earlier times, the no-island extremum gives the minimum and the radiation entropy grows with time as Hawking's original work indicates. 
However, as it turned out, at later times after the so-called Page time by which a significant portion of the black hole entropy is lost, a nontrivial extremum becomes the minimum and the radiation entropy starts following the thermodynamic entropy of the black hole. So the entire history of the radiation entropy yields the Page curve.

In this paper, we are concerned with the generalized entropy in the island formula \eqref{islandformula}. It consists of two parts: The purely geometric ``black hole'' entropy of the Bekenstein-Hawking type, ${\rm Area}(X)/(4G_N)$ \cite{Bekenstein:1972tm, Bekenstein:1973ur, Bardeen:1973gs, Hawking:1975vcx}, and the field-theoretic von Neumann (or fine-grained) entropy of the Hawking radiation, $S_{\rm vN}(\Sigma_X)$, in a fixed curved spacetime background. 
Instead of treating the geometric ``black hole'' and field-theoretic radiation parts separately, we will show that the generalized entropy for radiating black holes can be reproduced as a Wald-like entropy \cite{Wald:1993nt, Jacobson:1993vj, Iyer:1994ys} in a more unified geometric manner. 
The key idea is to directly work with the gravitational action that incorporates the backreaction of Hawking radiation as a conformal anomaly \cite{Christensen:1977jc, Callan:1992rs}.\footnote{This might only work for spacetimes in even dimensions which admit the local form of the conformal anomaly in terms of geometric invariants (see \cite{Duff:1993wm} and references therein). We will discuss how this idea can be generalized to four dimensions in section \ref{sec:Discussions}.}
It is then natural to identify the generalized entropy $S_{\rm gen}(X, \Sigma_X)$ with the Wald-like entropy of the gravitational theory coupled to the conformal anomaly (and conformal matter). 
This way, we obtain the generalized entropy in one go as a genuine geometric quantity without any non-gravitational field-theoretic input for the von Neumann entropy of Hawking radiation $S_{\rm vN}(\Sigma_X)$. 
It is worth emphasizing that our derivation is inherently Lorentzian without the need for Euclidean continuation.
Note that in this ``derivation'', the spacetime is everywhere gravitating including the radiation region $R$ outside the surface $D$ of the observer collecting  Hawking radiation. Put differently,  there is no region in spacetime where the background is fixed and gravitons cease to propagate. The diffeomorphism invariance must be imposed everywhere in spacetime, and when it is applied to higher dimensions, gravitons remain massless. So our findings touch upon the issues of the island proposal raised in the work of massive islands and inconsistency of islands in theories of long-range gravity \cite{Geng:2020qvw,Geng:2020fxl,Geng:2021hlu, Laddha:2020kvp, Raju:2021lwh, Geng:2021mic, Geng:2023qwm}.
We will expand on this point later wherever the appropriate context arises.

This paper is organized as follows. In section \ref{sec:Wald}, we briefly review the Wald entropy  and propose a generalization, referred to as Wald-like entropy, that is prescribed for the island formula.
In section \ref{sec:2d}, we illustrate our idea in detail and test the proposed Wald-like entropy by studying the black holes in two-dimensional dilaton-gravity theories, that is, the Jackiw-Teitelboim gravity \cite{Jackiw:1984je,Teitelboim:1983ux} and Russo-Susskind-Thorlacius model \cite{Russo:1992ax} as examples. 
After a short summary of the paper, a good part of section \ref{sec:Discussions} is devoted to the discussion on how a similar idea can be applied to higher dimensional black holes, focusing mostly on  the four-dimensional case.

\section{The Wald-like entropy for islands}
\label{sec:Wald}

In general relativity, there exist a family of Noether currents $J^a$ associated with the diffeomorphism invariance generated by an arbitrary vector $\xi^a$. Since the Noether current is conserved $\nabla_aJ^a=0$, the current $J^a$ can be locally expressed as $J^a=\nabla_bQ^{ab}$ where $Q^{ab}=-Q^{ba}$ is called Noether potential.
With some choice of $\xi^a$, we define the Wald-like entropy $S_W$ as the integral of a Noether potential $Q^{ab}$ over a codimension-two surface $X$:
\begin{align}\label{Waldlike}
S_W=2\pi\int_XQ^{ab}[\xi^c, \nabla^c\xi^d,\cdots]\epsilon_{ab}d\Omega_{D-2}
\end{align} 
where $d\Omega_{D-2}$ is the volume-form on $X$ and $\epsilon_{ab}$ is binormal to $X$ with the normalization $\epsilon_{ab}\epsilon^{ab}=-2$. It is implied that in the most general case, $Q^{ab}$ can, in principle, depend on $\xi^a$ and all its derivatives $\nabla^{b_1}\cdots\nabla^{b_n}\xi^a$. 
This is a generalization of the Wald entropy \cite{Wald:1993nt,Jacobson:1993vj, Iyer:1994ys}. In Wald's case, $X$ is a spacelike cross-section of the event horizon $H$ for stationary black holes and $\xi^a$ is the horizon-generating Killing vector. For Killing vectors, $Q^{ab}$ depends only on $\xi^a$ and its first-derivative  $\nabla^b\xi^a$, and the Wald entropy is independent of the choice of the cross-section $X$ and converges to the value evaluated at the bifurcation surface on which $\xi_a=0$ and $\nabla_{[a}\xi_{b]}=\kappa\epsilon_{ab}$ with $\kappa$ being the surface gravity. For example, in the case of two-derivative gravity theories with the Lagrangian density $L$, the Noether potential is given by $Q^{ab}=-{\delta L\over \delta R_{abcd}}\epsilon_{cd}$ and in particular, for the Einstein gravity, the Wald entropy yields the Bekenstein-Hawking entropy \cite{Wald:1993nt}.

When we try to apply Wald's idea to the ``derivation'' of the generalized entropy $S_{\rm gen}$ for the island formula \eqref{islandformula}, we encounter two immediate issues: (1) The extremal surface $X$ is {\it not} a cross-section of the Killing (event) horizon $H$. (2)  There is no timelike or null Killing vector in (evaporating) dynamical black holes. The first issue is present even in the case of islands for the eternal black hole \cite{Almheiri:2019yqk} in which the horizon-generating Killing vector exists.\footnote{To be more precise, even though the eternal black hole has the time translation isometry, prior to the extremization of the generalized entropy, the von Neumann entropy of Hawking radiation $S_{\rm vN}(\Sigma_X)$, being time-dependent, breaks the isometry. However, the maximization in time restores the time translation invariance.} The second issue arises generically for non-stationary black holes, regardless of islands, as discussed in \cite{Iyer:1994ys, Jacobson:1993vj}.

We propose the following prescription which, in disguise, looks identical to the one suggested in Wald's original paper \cite{Wald:1993nt} but differs from it in that the surface $X$ is not a cross-section of the horizon of the (dynamical) black hole but rather an extremal surface:
\begin{align}\label{genWaldlike}
S_{\rm gen}(X, \Sigma_X)=2\pi\int_XQ^{ab}[\xi^c, \nabla^c\xi^d]\epsilon_{ab}d\Omega_{D-2}\biggr|_{\xi^c\to 0, \nabla^{[c}\xi^{d]}\to \epsilon^{cd}}\ ,
\end{align}
where $\xi^a\to 0$ and $\nabla^{[c}\xi^{d]}\to \epsilon^{cd}$ by a suitable choice of the non-Killing vector $\xi^a$ which we will discuss momentarily. 
It should, however, be stated that as we only checked this proposal for the 2d dilaton gravity models, albeit including a dynamical evaporating black hole, some improvements might be necessary for the study of higher-dimensional (evaporating) black holes. 
We note in passing that in a study of the Wald entropy in the RST model \cite{Myers:1994sg}, a similar prescription was implicitly assumed and it was shown that, with $X$ being either a cross-section of the event or apparent horizon, the second law of thermodynamics holds for an evaporating black hole upon the inclusion of the entropy of Hawking radiation. 
In our proposal, $X$ is neither a cross-section of the event horizon nor an apparent horizon, and our claim and findings below also refine the analyses of \cite{Myers:1994sg} in that we examine the conformal anomaly (Hawking radiation) more carefully and reveal the connection of the Wald-like entropy to the generalized entropy in the island formula.\footnote{\label{footnote4}After the submission to the arXiv, we were informed by the authors of an earlier work \cite{Pedraza:2021cvx} that a similar idea was proposed and they made the observation that the 2d conformal anomaly reproduces the von Neumann entropy of Hawking radiation in a JT gravity example. In their follow-up work \cite{Pedraza:2021ssc}, they also found that for the 2d eternal black holes, the on-shell action in microcanonical ensemble was shown to be  minus the Wald(-like) entropy. Our expositions, takes, associated motivations and the details on this idea are different from theirs, besides working out the RST black hole examples, and in some sense, this work complements their proposal.}

This proposal assumes the existence of the (non-Killing) vector $\xi^a$ that has the properties $\xi^a=0$ and $\nabla^{[a}\xi^{b]}=\epsilon^{ab}$ on the extremal surface $X$.
There indeed exists such a vector:\footnote{This is somewhat similar to the use of the Kodama time \cite{Kodama:1979vn} in the black hole thermodynamics \cite{Hayward:1997jp} and Wald-like entropy \cite{Hayward:1998ee}. As in the case of the Kodama vector, $\xi^a$ itself is a conserved current $\nabla_a\xi^a=0$.}
\begin{align}\label{nonKillingvec}
\xi^a ={\cal N}\epsilon^{ab}\nabla_bS_{\rm gen}(x)\qquad\mbox{with}\qquad S_{\rm gen}(x)=S_{\rm gen}(X,\Sigma_X)
\end{align}
where $x$ denotes the two-dimensional coordinates that specify the location of the extremal surface $X$, and the normalization function ${\cal N}$ is adjusted such that $\nabla^{[a}\xi^{b]}=\epsilon^{ab}$ on $X$.
To elaborate on it, $\xi^a$ is obviously null and vanishes on $X$ by the definition of the extremal surface $X$. In addition, $\xi^a$ is normal to $X$ since $\epsilon^{ab}$ is binormal to the hypersurface $X$. We can then express $\nabla^{[a}\xi^{b]}=\kappa\epsilon^{ab}+t^{[a}\xi^{b]}$ where $t^a$ is tangential to $X$ and $\kappa$ would be the surface gravity in the case of stationary black holes \cite{Jacobson:1993vj}. Now, since $\xi^a=0$ on $X$, there exists a choice of ${\cal N}$ which yields $\nabla^{[a}\xi^{b]}=\epsilon^{ab}$ on $X$.
With this choice of the vector $\xi^a$, at least for the gravitational theories in which $Q^{ab}$ does not depend on higher derivatives of $\xi^a$, the generalized entropy is essentially prescribed to be the form identical to \eqref{genWaldlike}. Moreover, the extremization with respect to $X$ is a requirement on $\xi^a$ and built into this definition of the generalized entropy.
It may look self-fulfilling since we used what to be defined, $S_{\rm gen}(x)$, to define itself in the form we want. However, this prescription does not, a priori, require the explicit form of $S_{\rm gen}(x)$ and the a posteriori existence of the extremal surface so defined suffices for $S_{\rm gen}(X,\Sigma_X)$ to be well-defined and self-consistent.

The proposed generalized entropy \eqref{genWaldlike}, supplemented with \eqref{nonKillingvec}, for the island formula would not seem to satisfy the first law of thermodynamics since the proof of the first law relies on the vector $\xi^a$ being a Killing vector \cite{Wald:1993nt, Iyer:1994ys}. However, since it yields the Page curve \cite{Page:1993wv, Page:2013dx} which, after the Page time, {\it approximately} follows the thermodynamic entropy of the black hole \cite{Penington:2019npb,Almheiri:2019psf, Almheiri:2019hni}, it still needs to {\it approximately} obey the first law, at least, in the case of quasi-stationary black holes. Empirically speaking, given the resulting Page curves in some examples, that actually seems to be the case, but we do not have proof based on \eqref{genWaldlike} and \eqref{nonKillingvec}.  We leave this issue for the future. 

\section{Two-dimensional dilaton-gravity theories}
\label{sec:2d}

As an illustration of the above proposal, we study the black holes in two solvable models of two-dimensional dilaton-gravity theories: the Jackiw-Teitelboim (JT) gravity \cite{Jackiw:1984je,Teitelboim:1983ux} and the Russo-Susskind-Thorlacius (RST) model \cite{Russo:1992ax}, built on the Callan-Giddings-Harvey-Strominger (CGHS) model \cite{Callan:1992rs}, in which Hawking radiation is represented by conformal scalars. They are defined by the bulk action
\begin{equation}\label{gravityaction}
\begin{aligned}
I_G&=\int d^2x\sqrt{-g}\biggl[U(\phi)R+K(\phi)(4\lambda^2+\zeta g^{ab}\partial_a\phi\partial_b\phi)\\
&\qquad\qquad\qquad -{1\over 32\pi G}g^{ab}\partial_a\vec{f}\cdot\partial_b\vec{f}
-\underbrace{{c\over 24\pi}(g^{ab}\partial_a\chi\partial_b\chi + \chi R)}_{\rm conformal\,\,anomaly}\biggr]
\end{aligned}
\end{equation}
where $\phi$ is the dilaton, $\vec{f}$ denotes a collection of free massless scalar fields with the total central charge $c$, and $\chi$ is an auxiliary field (``conformalon''), satisfying $2\Box\chi=R$, which was introduced to render the Polyakov's non-local anomaly action \cite{Polyakov:1981rd} local.
The JT gravity and the RST model correspond to the cases
\begin{equation}\label{DefModels}
\begin{aligned}
\left(U(\phi), K(\phi), \lambda, \zeta\right)&=\left\{
\begin{array}{cc}
\left({\phi\over 16\pi G},{\phi\over 16\pi G}, {1\over\sqrt{2}}, 0\right) &\quad \mbox{(JT gravity)}\\
\left({e^{-2\phi}\over 16\pi G}-{c\phi\over 48\pi}, {e^{-2\phi}\over 16\pi G}, 1, 4\right) &\quad \mbox{(RST model)}
\end{array}
\right.\ .
\end{aligned}
\end{equation}
As mentioned earlier, the Wald(-like) entropy in the RST model was studied some time ago in \cite{Myers:1994sg}, and a more recent work \cite{Almheiri:2014cka} implicitly assumed the Wald(-like) entropy as the black hole entropy in these models. The Noether potential can be found as \cite{Iyer:1994ys}
\begin{align}
Q^{ab}=-\left(U(\phi)-{c\over 24\pi}\chi\right)\nabla^{[a}\xi^{b]}-2\xi^{[a}\nabla^{b]}\left(U(\phi)-{c\over 24\pi}\chi\right)\ .
\end{align}
Note that this final form only comes from the curvature-dependent terms in the action. All the other details of the model are eliminated by the use of the equations of motion.
Our proposal \eqref{genWaldlike} or the choice of $\xi^a$ in \eqref{nonKillingvec} then yields
\begin{align} \label{2dWaldlike}
S_{\rm gen}(X, \Sigma_X)=4\pi U(\phi)-{c\over 6}\chi\ ,
\end{align}
where $X$ is the extremal surface and we used $\epsilon^{ab}\epsilon_{ab}=-2$.

We are now going to demonstrate that our prescribed Wald-like entropy \eqref{2dWaldlike} correctly reproduces the generalized entropy for the island formula that agrees with the results  in \cite{Almheiri:2019hni,Almheiri:2019yqk, Almheiri:2019qdq, Gautason:2020tmk,Hartman:2020swn}.\footnote{In a more recent paper \cite{Yu:2022xlh}, the islands were studied in the Fabbri-Russo model \cite{Fabbri:1995bz} which is a one-parameter generalization of the RST model.}
A point we wish to make is that the field-theoretic von Neumann entropy of Hawking radiation $S_{\rm vN}(\Sigma_X)$ emerges from a unique solution of $\chi$ simply by solving the gravity equations of motion with the choice of an appropriate vacuum. It requires no information about the field-theoretic entanglement entropy. 
It should be mentioned that a similar observation was made in an earlier work \cite{Pedraza:2021cvx} as commented in footnote \ref{footnote4}.

\subsection{The JT gravity}
\label{sec:JT}

The JT gravity \cite{Jackiw:1984je,Teitelboim:1983ux} with a negative cosmological constant has an $AdS_2$ black hole solution and it is one of the first models studied in the development of the island idea \cite{Almheiri:2019psf, Almheiri:2019hni,Almheiri:2019yqk, Penington:2019kki, Almheiri:2019qdq}. So we first discuss the Wald-like entropy for the (eternal) $AdS_2$ black hole in the JT gravity:
\begin{align}\label{AdS2BH}
ds^2=-\left({2\pi\over\beta}\right)^2{dx^+dx^-\over\sinh^2\left({\pi\over\beta}(x^+-x^-)\right)}\equiv -e^{2\rho}dx^+dx^-\qquad\mbox{where}\qquad
x^{\pm}=\tau\pm\sigma\ ,
\end{align}
where we choose the range of $\sigma\in [-\infty, 0]$.
This is a solution to the equations of motion following from the action \eqref{gravityaction} with the choice of the set of functionals and parameters given in \eqref{DefModels}. In the conformal gauge defined by \eqref{AdS2BH}, the equations of motion are given by 
\begin{equation}\label{JTEOM}
\begin{aligned}
4\del_+\del_-\rho+e^{2\rho}&=0\ ,\\
\del_+\del_-\phi-2\phi\del_+\del_-\rho&={2G c\over 3} \del_+\del_-\chi\ ,\\
\del_+^2\phi-2\del_+\rho\del_+\phi&=-{2G c\over 3}\left(\partial_{+}\chi\partial_{+}\chi-\partial_{+}^2\chi+2\del_+\rho\del_+\chi\right)-{1\over 2}\del_+\vec{f}\cdot\del_+\vec{f}\ ,\\
\del_-^2\phi-2\del_-\rho\del_-\phi&=-{2G c\over 3}\left(\partial_{-}\chi\partial_{-}\chi-\partial_{-}^2\chi+2\del_-\rho\del_-\chi\right)-{1\over 2}\del_-\vec{f}\cdot\del_-\vec{f}\ ,\\
\del_+\del_-(\rho+\chi)&=0\ ,\\
\del_+\del_-\vec{f}&=0\ .
\end{aligned}
\end{equation}
The first is the $\phi$ equation of motion, $R+2=0$. The second to forth are the Einstein equations. The fifth and sixth are the $\chi$ and matter equations of motion, respectively.
The solution of our interest is the $AdS_2$ black hole \eqref{AdS2BH} with the static dilaton without classical matter \cite{Almheiri:2014cka}:
\begin{align}
\phi=-{2\pi\phi_{\rm r}\over\beta}\coth{{\pi\over\beta}(x^+-x^-)}+{cG\over 3}\ ,\qquad\qquad\vec{f}=0\ .
\end{align}
The most important in this discussion is the ``conformalon'' solution
\begin{align}
\chi = -\rho +h_+(x^+) + h_-(x^-)\ ,
\end{align}
where the homogeneous part of the solution denoted by the chiral functions $h_{\pm}$ satisfy
\begin{align}
\partial_{\pm}h_{\pm}\partial_{\pm}h_{\pm}-\partial_{\pm}^2h_{\pm}=\partial_{\pm}\rho\partial_{\pm}\rho-\partial_{\pm}^2\rho=\left({\pi\over\beta}\right)^2\ .
\end{align}
Note that this corresponds to the Hartle-Hawking state as the choice of vacuum with the thermal radiation in the Rindler coordinates. 
These equations follow from the fact that the LHS of the third and fourth equations vanish with the above choice of $\rho$ and $\phi$.
The general solution can be expressed as\footnote{The solution in \cite{Almheiri:2014cka} corresponds, in our notation, to $h_{\pm}=\pm{\pi\over\beta}x^{\pm}$ which is a limiting solution with the choice, $c^{\pm}=\Lambda+\ln(\beta/(4\pi))$ and $x_0^{\pm}=\pm\Lambda$ with $\Lambda\to +\infty$.}
\begin{align}
h_{\pm}(x^{\pm})=c^{\pm}-\ln\left(\mp{\beta\over 2\pi}\sinh\left({\pi\over\beta}(x^{\pm}-x_0^{\pm}\right)\right)
\end{align}
with the integration constants $c^{\pm}$ and $x_0^{\pm}$.

Having specified the solution to the gravity equations of motion \eqref{JTEOM}, we can now calculate the Wald-like entropy \eqref{2dWaldlike} as a function of the location $(x^+, x^-)$ of the extremal surface $X$:
\begin{align}\label{AdSgenEntropy}
S_{\rm gen}=s_0+{\pi\phi_{\rm r}\over 2G\beta}\coth{\pi(x^--x^+)\over\beta}
+{c\over 6}\ln{\beta\sinh{{\pi\over\beta}(-x^++x_0^+)}\sinh{{\pi\over\beta}(x^--x_0^-)}\over 2\pi\sinh{\pi(x^--x^+)\over\beta}}
\end{align}
where $s_0={c\over 12}-{c\over 6}(c^++c^-)$ which is an ambiguity in the additive constant \cite{Witten:2021unn}. 
The extremization with respect to $x^{\pm}$ determines the position $(x^+, x^-)$ of the extremal surface in terms of $(x_0^+, x_0^-)$.  
This indeed reproduces the results in \cite{Almheiri:2019psf, Almheiri:2019hni,Almheiri:2019yqk, Almheiri:2019qdq}, up to an additive constant, after the maximization in time $\tau$ \cite{Hubeny:2007xt} which sets $\tau=\tau_0$ eliminating the time-dependence. Note that there are no UV cutoffs appearing in the radiation (anomaly) part of the entropy in contrast to the field-theoretic computation and this result does not require regularization or renormalization.\footnote{It is tempting to interpret this finiteness of the gravitational generalized entropy as a realization of  the idea by Susskind and Uglum \cite{Susskind:1994sm} as recently emphasized by Witten in the context of von Neumann algebra \cite{Witten:2021unn}. }

The spacelike region $\Sigma_X$ in \eqref{islandformula} corresponds to the interval $\Sigma_X=B_X\in [\sigma, \sigma_0]$, or equivalently, its complement $I_X\cup R$ as defined below \eqref{islandformula}. From the viewpoint of our derivation, it is natural to interpret the position $(x_0^+, x_0^-)$ as that of an observer collecting Hawking radiation {\it within} the $AdS_2$ boundary, i.e., $\sigma_0<0$ in our convention, since our spacetime is nothing other than the $AdS_2$ black hole. However, in the setup of the islands \cite{Almheiri:2019psf, Almheiri:2019hni,Almheiri:2019yqk, Almheiri:2019qdq}, the location $\sigma_0$ of the boundary of $\Sigma_X$ is placed, beyond the $AdS_2$ boundary ($\sigma_0\to 0_-$), in the external bath regions ($\sigma_0>0$) with a transparent interface condition so that the radiation is transmitted across the $AdS_2$ boundary rather than reflected back in.  Nevertheless, we point out that for the eternal black hole that does not evaporate, the islands (outside the horizon) actually exist for $\sigma_0<0$, that is, even when the observer at the boundary of $\Sigma_X$ is placed within the $AdS_2$ boundary, hence yielding a Page curve. Since $\sigma_0$ is a free parameter, we can choose it to be positive as one pleases, but it is unclear from this perspective whether it should be interpreted as an observer in the external bath attached to the $AdS_2$ black hole.
At this point, it is worth emphasizing that the computation is fully gravitational and we made no reference to the non-gravitating regions to derive the generalized entropy associated with the island formula. 
However,  it then raises the issue of diffeomorphism invariance. We took it for granted that we can associate a physical meaning to the position $(x_0^+, x_0^-)$, but in a diffeomorphism invariant theory, that may not be the case without explicitly breaking diffeomorphism invariance.  So it seems that we cannot factorize the Hilbert space into three regions, $I_X$, $R$, and $B_X$. We then lose the interpretation of the Wald-like entropy as the fine-grained entropy of Hawking radiation  \cite{Geng:2020qvw,Geng:2020fxl,Geng:2021hlu, Laddha:2020kvp, Raju:2021lwh, Geng:2021mic, Geng:2023qwm}.

To be self-contained, we now turn to the no-island case which appears in the two-sided eternal black hole. The entropy must be given by
\begin{align}\label{AdSnoisland}
S_{\rm no-island}={c\over 6}\ln{\beta\sinh{{\pi\over\beta}(-x^++x_0^+)}\sinh{{\pi\over\beta}(x^--x_0^-)}\over 2\pi\sinh{\pi(x^--x^+)\over\beta}}
=-{c\over 6}\chi(x^+,x^-)\biggr|_{x^{\pm}=- x_0^{\mp}- i\beta/2}
\end{align}
where $x^{\pm}$ is located in the left Rindler wedge at the mirror point of $x_0^{\pm}$ and in terms of the Kruskal coordinates, $y^{\pm}=\pm e^{\pm{2\pi\over\beta}x^{\pm}}=\mp e^{\mp{2\pi\over\beta}x_0^{\mp}}=y_0^{\mp}$. The no-island entropy grows linearly in the Rindler time $\tau$ at late times.
It seems that there is no first-principle derivation of the no-island ``extremum'' from the Wald-like entropy \eqref{2dWaldlike}.
Naively, if $X=\O$, the Wald-like entropy \eqref{2dWaldlike} would have to vanish. One of the issues is that the geometric part ${\rm Area}(X)/(4G_N)$ and field-theoretic part $S_{\rm vN}(\Sigma_X)$ in \eqref{islandformula} are unified into a single geometric entity in the Wald-like entropy \eqref{2dWaldlike}. So, conceptually speaking, it is hard to separate them and single out $S_{\rm vN}(\Sigma_X)$ from some first principle. However, it is fairly clear that the radiation entropy for the empty set $X=\O$ is, by definition, extracted as
\begin{align}\label{ournoisland}
S_{\rm no-island}=-{c\over 6}\chi(x^+,x^-)|_{y^{\pm}=y_0^{\mp}\,\,\left(x^{\pm}=- x_0^{\mp}- i\beta/2\right)}\ .
\end{align}
So practically, we can adopt this formula as our prescription for the no-island radiation entropy.

To summarize the technical part of this discussion on the JT gravity example, the minimum of the Wald-like entropy \eqref{AdSgenEntropy} and no-island radiation entropy \eqref{ournoisland} reproduces the Page curve in 
\cite{Almheiri:2019hni,Almheiri:2019yqk, Almheiri:2019qdq}.

\subsection{The RST model}
\label{sec:RST}

In the JT gravity we have just seen that the generalized entropy for the island formula can be derived without reference to the external non-gravitating bath attached to the $AdS_2$ black hole. Moving forward, in the case of asymptotically flat black holes, even though there is no external bath attached in any way, the island formula still assumes the non-gravitating (radiation) region $R$ because the entanglement entropy of Hawking radiation is the field-theoretic von Neumann entropy in a fixed curved spacetime background where gravity is treated as non-dynamical \cite{Almheiri:2020cfm}.
Having this point in mind, in this section, our aim is to demonstrate that the generalized entropy can be derived as the Wald-like entropy without assuming the non-gravitating region by studying the asymptotically flat black holes in the RST model \cite{Russo:1992ax}.
The generalized entropy in the RST model was first studied in \cite{Fiola:1994ir} long time ago. More recently, it was reinvestigated in light of the island proposal \cite{Gautason:2020tmk, Hartman:2020swn}. As we will see, the Wald-like entropy \eqref{2dWaldlike} reproduces the results in \cite{Fiola:1994ir, Gautason:2020tmk, Hartman:2020swn}.\footnote{To be more precise, if we understand correctly, the results in \cite{Gautason:2020tmk} differ from those in \cite{Fiola:1994ir, Hartman:2020swn}, essentially, by ${c\over 6}\rho(x^+,x^-)$. The difference is traced back to the sign of the linear $\phi$ term in $U(\phi)$ given in \eqref{DefModels}. Our results agree with the latter.}

In addition to the diffeomorphism invariance, the RST model has an extra symmetry associated with the current conservation $\nabla^a\left(\nabla_a(\phi-\rho)\right)=0$. By virtue of this symmetry, working in the conformal gauge 
\begin{align}
ds^2=-e^{2\rho}dx^+dx^-\ , 
\end{align}
we can consistently set 
\begin{align}\label{Kruskalgauge}
\phi=\rho\ .
\end{align}
Then the equations of motion that follow from the action \eqref{gravityaction} with the choice \eqref{DefModels} can be expressed as\footnote{For the RST model, we work in the unit $8G=1$ as it is conventional in the literature.}
\begin{equation}\label{RSTEoM}
\begin{aligned}
\partial_+\partial_-\Omega +1 &= 0\ , \\
-\partial_{\pm}^2\Omega &= {c\over 12}\biggl(\left(\partial_{\pm}(\chi+\rho)\right)^2-\partial^2_{\pm}(\chi+\rho)\biggr)+{1\over 2}\partial_{\pm}\vec{f}\cdot \partial_{\pm}\vec{f}\ ,\\
 \partial_+\partial_-(\chi+\rho)&=0\ ,
\end{aligned}
\end{equation}
where  a convenient field redefinition was introduced \cite{Bilal:1992kv, deAlwis:1992emy}
\begin{align}
\Omega = e^{-2\phi}+{c\over 24}\phi\ .
\end{align}
There are exact black hole solutions to these equations of motion \cite{Russo:1992ax}: (1) eternal black hole and (2) evaporating black hole.

We first discuss the eternal black hole which is given by 
\begin{align}\label{RSTBH}
\Omega = M-x^+x^-\ge \Omega_c\qquad\mbox{with}\qquad\Omega_c={c\over 48}\left(1-\ln{c\over 48}\right)\ ,
\end{align}
where $\Omega_c$ corresponds to the critical value of the dilaton, $\phi_c=-{1\over 2}\ln(c/48)>\phi$. This is the value at which the determinant of the $(\rho, \phi)$-kinetic terms vanishes and can be thought of as a strong coupling singularity where the spacetime ends.
The matter $\vec{f}=0$ and the ``conformalon'' $\chi$ is given by 
\begin{align}
\chi = -\rho +h_+(x^+)+h_-(x^-)\qquad\mbox{with}\qquad \partial_{\pm}h_{\pm}\partial_{\pm}h_{\pm}-\partial_{\pm}^2h_{\pm}=0\ .
\end{align}
Note that this corresponds to the Hartle-Hawking state as the choice of vacuum with no radiation in the Kruskal coordinates. 
The chiral functions $h_{\pm}(x^{\pm})$ can be found as
\begin{align}\label{Kruskalh}
h_{\pm}(x^{\pm})=c_{\pm}-\ln\left(\pm(x^{\pm}-x_0^{\pm})\right)\ .
\end{align}
For the two-sided black hole, the Wald-like entropy \eqref{2dWaldlike} is thus found to be
\begin{align}
S_{\rm gen}=2s_0+4\Omega(x^+,x^-)+{c\over 6}\ln(x^+-x_0^+)^2(x^--x_0^-)^2
\end{align}
where we used that $4\pi U(\phi)+{c\over 6}\rho=2\Omega$ in the unit $8G=1$ and $s_0=-{c\over 6}(c^++c^-)$ is an additive constant ambiguity.
The extremization with respect to $x^{\pm}$ determines the position $(x^+, x^-)$ of the extremal surface in terms of the position $(x_0^+, x_0^-)$ of an observer collecting Hawking radiation.
This agrees with \cite{Hartman:2020swn} up to a constant which does not affect the location of the extremal surface. We note again that there are no UV cutoffs appearing in the radiation part of the entropy in contrast to the field-theoretic computation. Most importantly, once again, we made no reference to the non-gravitating regions in this derivation.

With the prescription for the no-island radiation entropy in \eqref{ournoisland}, since the $x^{\pm}$ coordinates are Kruskal as implied in  \eqref{RSTBH}, we find that
\begin{align}
S_{\rm no-island}={c\over 6}\left(\ln(x_0^+-x_0^-)^2+\rho(x_0^-, x_0^+)\right)\ .
\end{align}
At late times, this grows linearly in the Minkowski time $\tau$ defined via the map $x^{\pm}=\pm e^{\pm(\tau\pm\sigma)}$ and agrees with the results in \cite{Gautason:2020tmk, Hartman:2020swn}.

We next discuss the evaporating black hole created by a shockwave $T^f_{++}={M\over x_s^+}\delta(x^+-x_s^+)$ at $x^+=x_s^+$. It is described by
\begin{equation}
\begin{aligned}\label{evaporatingBH}
\Omega  &= -{M\over x_s^+}(x^+-x_s^+)\Theta(x^+-x_s^+)-x^+x^- -{c\over 48}\ln(-x^+x^-)\ge \Omega_c\ ,\\
\chi &= -\rho +h_+(x^+) + h_-(x^-)\ ,
\end{aligned}
\end{equation}
where $\Theta(x^+-x_s^+)$ is a step function and $h_{\pm}$ will be determined shortly. Before the shockwave is shot in, the spacetime is in the so-called linear dilaton vacuum since $\phi=\rho=-{1\over 2}\ln(-x^+x^-)=-\sigma$ for $x^+<x_s^+$. The metric in the linear dilaton vacuum is flat and given by
\begin{align}
ds_{LD}^2=-e^{2\rho}dx^+dx^-=-d(\ln x^+)d(-\ln(- x^-))\ .
\end{align}
In this setup, we want to study the radiation entropy in the Minkowski vacuum  \cite{Fiola:1994ir, Gautason:2020tmk, Hartman:2020swn}, meaning that the stress tensor and Liouville field $\rho$ are defined with respect to the Minkowski coordinates $(w^+, w^-)=(\ln x^+, -\ln(-x^-))$. After the shock $x^+>x_s^+$, they are given by
\begin{equation}
\begin{aligned}\label{trhotransform}
\left({\del_{\pm}w^{\pm}\over \del_{\pm}x^{\pm}}\right)^2\tilde{t}_{\pm}(w^{\pm})&= t_{\pm}(x^{\pm})+{c\over 24}\{w^{\pm},x^{\pm}\}=0\ ,\\
\tilde{\rho}(w^+,w^-)&=\rho(x^+,x^-)+{1\over 2}\ln\left({\del x^+\over \del w^+}{\del x^-\over \del w^-}\right)
=\rho(x^+,x^-)+{1\over 2}\ln(-x^+x^-)\ ,
\end{aligned}
\end{equation}
where $t_{\pm}\equiv {c\over 12}(\del_{\pm}h_{\pm}\del_{\pm}h_{\pm}-\del_{\pm}^2h_{\pm})$ and the Schwarzian derivative $\{w, x\}\equiv w'''/w'-3/2(w''/w')^2$.
The middle expression of the first equation vanishes by using the second equation of motion in \eqref{RSTEoM}.
The field $\Omega$ is a scalar  and so unchanged since it is a field redefinition of the dilaton.\footnote{To be precise, in the choice \eqref{Kruskalgauge}, the Liouville field $\rho$ is the one in the Kruskal coordinates $\rho(x^+,x^-)$. Since the dilaton $\phi$ is a scalar whereas the Liouville field $\rho$ is not, the coordinate system has to be specified in \eqref{Kruskalgauge} and the choice is the Kruskal coordinates $(x^+, x^-)$.} From the first equation in \eqref{trhotransform}, $\tilde{h}_{\pm}$ are fixed to be
\begin{align}
\tilde{h}_{\pm}(w^{\pm})=c_{\pm}-\ln\left(\pm(w^{\pm}-w_0^{\pm})\right)\ ,
\end{align}
where $\chi(w^+,w^-)=-\tilde{\rho}(w^+,w^-)+\tilde{h}_+(w^+)+\tilde{h}_-(w^-)$. This is the same in form as \eqref{Kruskalh}. However, it is important that this expression is in the Minkowski coordinates whereas \eqref{Kruskalh} is in the Kruskal coordinates. Accordingly, this corresponds to the Boulware state as the choice of vacuum with no radiation in the Minkowski coordinates.

Having specified the evaporating black hole solution, the Wald-like entropy \eqref{2dWaldlike} yields
\begin{align}\label{SgenRSTBH}
S_{\rm gen}=s_0+2\Omega(x^+,x^-)+{c\over 12}(w^+-w^-)+{c\over 6}\ln(w^+-w_0^+)(-w^-+w_0^-)\ ,
\end{align}
where we used again that $4\pi U(\phi)+{c\over 6}\rho=2\Omega$ in the unit $8G=1$ and $x^{\pm}=\pm e^{\pm w^{\pm}}$. As the observer is sent to ${\cal I}^+$, i.e., for a very large $x_0^+$ for which $w^+-w_0^+\approx -w_0^+$, this agrees with \cite{Hartman:2020swn} up to a constant that does not affect the location of the extremal surface.
The extremization with respect to $w^{\pm}$ determines the position $(w^+, w^-)$ of the extremal surface in terms of the position $(w_0^+, w_0^-)$ of an observer collecting Hawking radiation.

As in the JT gravity example, we emphasize, once again, that the computation is fully gravitational and we made no reference to the non-gravitating regions to derive the generalized entropy. In this case, the issue of diffeomorphism invariance is sharper because the asymptotically flat region (with a linear dilaton) is simply present without attaching an external bath and $(w_0^+, w_0^-)$ is unambiguously a point in the asymptotiically flat spacetime. 
The existence of the island depends on whether or not a legitimate physical meaning can be associated to the point $(w_0^+, w_0^-)$ even though we took it for granted.
In a diffeomorphism invariant theory, it is far from obvious and it usually is not the case simply because the point $(w_0^+, w_0^-)$ is not diffeomorphism invariant. 
So the Hilbert space does not factorize into three regions, $I_X$, $R$, and $B_X$. We then lose the interpretation of the Wald-like entropy as the fine-grained entropy of Hawking radiation  \cite{Geng:2020qvw,Geng:2020fxl,Geng:2021hlu, Laddha:2020kvp, Raju:2021lwh, Geng:2021mic, Geng:2023qwm}. One possible way-out would be to vary $(w_0^+, w_0^-)$ and minimize the generalized entropy $S_{\rm gen}$ with respect to both $w^{\pm}$ and $w_0^{\pm}$ \cite{Geng:2021mic, Geng:2023qwm}, but it then yields zero entropy since the Page curve after the Page time decreases to zero in the infinite future in retarded time.

To be self-contained, we now turn to the no-island case. We are interested in the growth of the radiation entropy with respect to the observer's Minkowski time after the shock $x^+>x_s^+$. So the coordinates of our interest are $(\tilde{w}^+, \tilde{w}^-)$ defined by $x^+=e^{\tilde{w}^+}$ and $M+x^-=-e^{-\tilde{w}^-}$, given the metric \eqref{evaporatingBH}, which are related to the Minkowski coordinates $(w^+, w^-)$ on ${\cal I}^-$ by $(w^+, w^-)=(\tilde{w}^+, \tilde{w}^- -\ln(1+Me^{\tilde{w}^-}))$.
Now, the evaporating black hole ends at $\Omega=\Omega_c$ which is timelike before and some time after the shockwave creates the black hole. 
The location of the endpoint $(w^+, w^-)$ depends on the boundary condition at $\Omega=\Omega_c$. 
We impose a reflection condition on the timelike boundary after the shock: $(\tilde{w}^+, \tilde{w}^-)=(\tilde{w}_0^-, \tilde{w}_0^+)$.\footnote{This is similar to but slightly different from the boundary condition in \cite{Fiola:1994ir, Hartman:2020swn}.}
Then the prescription \eqref{ournoisland} adjusted to this case reads
\begin{equation}
\begin{aligned}
S_{\rm no-island}&={c\over 6}\biggl[\ln(\tilde{w}_0^--\tilde{w}_0^+)\left(\tilde{w}_0^--\tilde{w}_0^+-{1\over 2}\ln{1+Me^{\tilde{w}_0^-}\over 1+Me^{\tilde{w}_0^+}}\right)+\rho(e^{\tilde{w}_0^-}, -M-e^{-\tilde{w}_0^+})\\
&\qquad +{1\over 2}(\tilde{w}_0^--\tilde{w}_0^+)+{1\over 2}\ln(1+Me^{\tilde{w}_0^+})\biggr]\ .
\end{aligned}
\end{equation}
At late times the second line dominates and the radiation entropy grows as $c\tilde{w}_0^-/12$ which agrees with  \cite{Gautason:2020tmk, Hartman:2020swn} even though the details are different. The minimization of the island and no-island entropies yields the Page curve as demonstrated in \cite{Gautason:2020tmk, Hartman:2020swn}.

To summarize the technical part of our discussion on the eternal and evaporating black holes in the two-dimensional dilaton-gravity models, we successfully reproduced the known generalized entropy for the island formula from our proposed Wald-like entropy \eqref{genWaldlike} with $\xi^a$ in \eqref{nonKillingvec}.

\section{Discussions}\label{sec:Discussions}

We proposed a {\it Lorentzian} derivation of the generalized entropy associated with the island formula as a Wald-like entropy. The derivation is {\it fully gravitational without reference to the exterior non-gravitating region or field-theoretic von Neumann entropy of Hawking radiation in a fixed curved spacetime background}. We illustrated how this idea actually works in detail by studying the radiating black holes in the two-dimensional dilaton-gravity theories. 
Importantly, our findings and those in \cite{Pedraza:2021cvx} touch upon the issues of the island proposal raised in the work of massive islands and inconsistency of islands in theories of long-range gravity \cite{Geng:2020qvw,Geng:2020fxl,Geng:2021hlu, Laddha:2020kvp, Raju:2021lwh, Geng:2021mic, Geng:2023qwm}. In the 2d case, in the absence of gravitons,  the issue is better phrased as that of diffeomorphism invariance. As we argued, they exacerbate the issues and obscure the foundation of the island proposal as a resolution to the information paradox. Namely, operationally, the islands seem to emerge and the Page curve follows without setting up a non-graviting region. However, without breaking diffeomorphism invariance by freezing gravity there, {\it i.e.}, giving a legitimate physical meaning (such as a putative observer) to $w_0^{\pm}$ in \eqref{SgenRSTBH} as a quantity in a theory of gravity, the Hilbert space does not factorize and the entropy so obtained loses its interpretation as the fine-grained entropy of Hawking radiation. 

In principle, a similar idea can be generalized to higher dimensions. However, since the conformal anomaly plays an important role, the applicability might be limited to even dimensions  in which the conformal anomaly can be expressed by local geometric invariants. Here, we would like to discuss the four-dimensional generalization. First, it is reasonable to assume that Hawking radiation is dominated by massless matter, in particular, conformally-coupled massless matter. So the backreaction of Hawking radiation is incorporated into the gravity theories as the conformally-coupled matter action and their conformal anomaly. In higher dimensions, there can be a non-anomaly contribution since, for example, the conformally-coupled scalars $\vec{f}$ have the explicit curvature coupling $R\vec{f}^2$ on which the Wald-like entropy can depend.\footnote{In contrast, the conformal anomaly is all that matters in two dimensions because the matter action does not explicitly depend on the curvature and after integrating them out, the only curvature-dependence comes from the conformal anomaly. We thank Sumit Das for discussions on this point.} With this remark in mind, recall that the homogeneous part of the ``conformalon'' solution to $2\Box\chi=R$ played very important role in our success in two dimensions. In four dimensions, a similar local form of the anomaly action that serves our purpose would be the one constructed by Mottola \cite{Mottola:2016mpl} which is a covariant version of the so-called Riegert action \cite{Riegert:1984kt} and quadratic in the conformalon $\chi$ similar to \eqref{gravityaction}.\footnote{An alternative local action is the one with the Nambu-Goldstone dilaton $\phi$  \cite{Schwimmer:2010za} which is quartic in $\phi$ and was used in a proof of the $a$-theorem \cite{Komargodski:2011vj}.} Mottola's action contains the $a$ and $c$ anomalies, and if needed, we can add the scheme-dependent $b$ anomaly contribution which is given by a local $R^2$ action.
It is then straightforward to run a similar program to that we carried out in two dimensions, but the difficulty will be to solve the equations of motion for the gravity $+$ conformal matter $+$ anomaly system. However challenging it may be, the test is to see if the Wald-like entropy reproduces the salient features of the generalized entropy for four-dimensional black holes found in earlier works \cite{Almheiri:2019psy,Hashimoto:2020cas,Chen:2020hmv,He:2021mst}.
It is worth emphasizing that since it is a fully gravitational computation, the spacetime is everywhere gravitating including the radiation region $R$ outside the surface of an observer collecting Hawking radiation. So gravitons remain massless and the findings in higher dimensions would further corroborate the aforementioned issues raised in   \cite{Geng:2020qvw,Geng:2020fxl,Geng:2021hlu, Laddha:2020kvp, Raju:2021lwh, Geng:2021mic, Geng:2023qwm}.

A natural question is whether and how our Lorentzian derivation is related to the replica wormholes in the Euclidean path integral \cite{Penington:2019kki,Almheiri:2019qdq}. At the moment, we do not have much to say about it except that the Wald-like entropy must correspond to the wormhole saddle, or rather, it is hard to see the ``Hawking no-island'' saddle in our approach. As commented in footnote \ref{footnote4}, it is worth mentioning that the works \cite{Pedraza:2021ssc, Svesko:2022txo} address part of this issue extending an earlier work \cite{Iyer:1995kg}. Namely, using the Euclidean path integral, they showed that  the on-shell action in microcanonical ensemble was shown to be  minus the Wald(-like) entropy, connecting the Lorentzian derivation and Euclidean path integral, in the case of the $AdS_2$ black hole and $dS_2$. However, their connection bypasses the replica wormholes.
On that score, it might be useful to study the issue of  Lorentzian vs. Euclidean in a more algebraic 2d CFT description of the replica wormholes \cite{Hirano:2021rzg}.


\section*{Acknowledgments}

SH would like to thank Sumit Das, Robert de Mello Koch, and Koji Hashimoto for discussions and the mathematics and physics departments at Nagoya University and the Department of Physics and the Yukawa Institute for Theoretical Physics at Kyoto University for their hospitalities during his visits where part of this work was done. The work of SH is supported in part by the National Natural Science Foundation of China under Grant No.12147219.


\end{document}